\documentclass[twocolumn,showpacs,preprintnumbers,amsmath,amssymb,floatfix]{
 revtex4}

\usepackage{graphicx}% Include figure files
\usepackage{dcolumn}% Align table columns on decimal point
\usepackage{bm}% bold math
\usepackage{amsmath}
\usepackage{sidecap}
\begin{document}

\preprint{APS/123-QED}

\title{The Dynamic Scaling Study of Vapor Deposition Polymerization: A Monte
Carlo Approach}

\author{Sairam Tangirala}
\email{sairam@hal.physast.uga.edu}
\author{D.P. Landau}
\affiliation{
Center for Simulational Physics, The University of Georgia, Athens, GA 30602}

\author{Y.-P. Zhao}
\affiliation{
Nanoscale Science and Engineering Center, Department of Physics and Astronomy,
The University of Georgia, Athens, GA 30602}

\date{\today}% It is always \today, today,

\begin{abstract}
The morphological scaling properties of linear polymer films grown by vapor
deposition polymerization (VDP) are studied by 1+1D Monte Carlo simulations. The
model implements the basic processes of random angle ballistic deposition
($F$), free-monomer diffusion ($D$) and monomer adsorption along with the
dynamical processes of polymer chain initiation, extension, and merger. The
ratio $G=D/F$ is found to have a strong influence on the polymer film
morphology. Spatial and temporal behavior of kinetic roughening has been
extensively studied using finite-length scaling and height-height correlations $H(r,t)$. The scaling
analysis has been performed within the no-overhang approximation and the scaling behaviors at local and global length scales were found to be very different. The global and local scaling exponents for morphological evolution have been evaluated for varying free-monomer diffusion by growing the films at $G$ = $10$,
$10^2$, $10^3$, and $10^4$ and fixing the deposition flux $F$. With an increase in $G$ from $10$ to $10^4$, the average growth
exponent $\beta \approx 0.50$ was found to be invariant, whereas the global roughness exponent $\alpha_{g}$ decreased from $0.87(1)$ to $0.73(1)$
along with a corresponding decrease in  the global dynamic exponent $z_g$ from $1.71(1)$ to $1.38(2)$. The global scaling exponents were
observed to follow the dynamic scaling hypothesis, $z_g=\alpha_{g}/\beta$. With a similar increase in  $G$ however, the average local roughness exponent ${\alpha_{l}}$ remained close to  $0.46$ and the anomalous growth exponent ${\beta}_{\ast}$ decreased from 0.23(4) to 0.18(8).  The interfaces display anomalous scaling and multiscaling in the relevant height-height correlations. The variation of $H(r,t)$ with deposition time $t$ indicates non-stationary growth. A comparison has been made between the simulational findings and the experiments wherever applicable.

\end{abstract}

\maketitle

\section{\label{sec:level1}INTRODUCTION}
Our motivation for gaining theoretical understanding of polymer thin film growth
stems from their technological applications in microelectronic interconnects
\cite{ART_LU_AND_MOORE,BOOK_WONG}, organic electronics \cite{BOOK_WONG}, and
biotechnology. Various experimental methods like vapor deposition polymerization
(VDP) \cite{ART_LU_AND_MOORE,Gorham,Lahann,szulczewski:1875}, ionization
assisted polymer deposition \cite{Usui}, sputtering growth
\cite{Biederman200327}, pulsed laser deposition \cite{BOOK_Chrisey,Piqu2003293},
and organic molecular beam deposition \cite{Schreiber} have been developed to
produce a variety of polymer thin films. Polymer film growth is complex compared
to the conventional inorganic thin film growth process due to polymer's
complicated structure and interactions that include internal degrees of freedom,
limited bonding sites, chain-chain interactions, etc. Many experimental efforts
have focussed on the formation of polymer thin films using VDP
\cite{PhysRevLett.73.708,PhysRevLett.78.2389,ART_FORTIN_1}. In a typical VDP
experiment, a wafer (2-D substrate) is exposed to one or more volatile gas phase
precursors that produce free-monomers. The free-monomers impinge on the
substrate at random locations and react on the substrate surface to produce the
desired deposit. Polymer thin films grown by VDP are made up of long polymer
chains formed through the polymerization reaction occurring during the growth
process. The polymerization process involves the interaction of two free-monomer
molecules in a chemical reaction to initiate a dimer (polymer chain of length =
2). The free-monomers moving towards the substrate are consumed by either of the
two processes: first being chain initiation in which new polymer molecules are
generated; and secondly, chain propagation in which the existing polymer
molecules are extended to higher molecular weight. Besides these two mechanisms,
the free-monomer adsorption, diffusion, and polymer merger can be considered as other mechanisms that determine the overall film
morphology. The chemical nature of the linear polymer chain restricts the number
of bonding sites. A free-monomer can only bond to either of the two active ends
of a polymer, or to another free-monomer. This bonding constraint leads to the
formation of  an entangled or an overhang configuration, which blocks the region
it covers from the access of other incoming free-monomers. In conventional
physical vapor deposition (PVD) processes
\cite{BOOK_Mattox}, atoms can nucleate at the nearest neighbors of the nucleated
sites and  atomistic processes such as surface diffusion, edge diffusion, step
barrier effect, etc. effect  the growth, resulting in the films being compact
and dense\cite{BOOK_BARABASI_STANLEY,BOOK_MEAKIN,PhysRevB.49.10597,
PhysRevE.58.7571}. Recent investigations by Zhao et al.\ \cite{PhysRevE.60.4310} have shown that the
submonolayer growth behavior of VDP is very different from that of PVD due to
long chain confinement and limited bonding sites, indicating that the detailed molecular configuration can drastically change the
growth behavior \cite{ART_BOWIE_ZHAO}.
\indent 
In experiments, the growth behavior of polymer thin films  have been investigated
through  their morphological evolution study.  The VDP processes
for producing Parylene-N (PA-N) films typically are far from equilibrium. The
precursor material di-p-xylylene (dimer) is sublimed at $150^\circ$C and then
pyrolized into free-monomers at $650^\circ$C. The free-monomers impinge at
random angles onto the Si-wafer at room temperature and eventually condense and
polymerize to form the polymer film.  By varying the growth rates
in the PA-N growth experiments, Zhao et al.\  \cite{ART_ZHAO_FORTIN_1} reported 
an average roughness exponent $\alpha=0.72\pm0.05$ and an average growth exponent
$\beta=0.25\pm0.03$. However, by considering the tip effect of the atomic force
microscope \cite{aue:1347}, the range of $\alpha$ was estimated to be between
$0.5$ to $0.7$ and the authors found the absence of dynamic scaling hypothesis in the PA-N film growth.
In the recent experiments done by Lee et al. \cite{lee:115427}, the authors observed unusual changes in the roughening behavior
during the poly(chloro-p-xylylene) growth. In the early rapid growth regime,
they observed  $\beta = 0.65$ (larger than the random
deposition $\beta = 0.5$) and upon complete
coverage of the substrate (around $d=10 nm$), they found $\beta =
0.0$ and the interface width did not evolve with the film thickness. Finally,
during the continuous growth regime, the surface roughness again was found to increase
steadily with a new power law of $\beta = 0.18$  \cite{lee:115427}, which is
close to the results of Zhao et al. \cite{ART_ZHAO_FORTIN_1}. Of the known theoretical results for dynamic roughening, the MBE nonlinear
surface diffusion dynamics proposed by Lai and Das Sarma
\cite{PhysRevLett.66.2348} predicted similar exponents as obtained in the
experimental study of Zhao. et al. \cite{ART_ZHAO_FORTIN_1}. However, their
nonlinear surface diffusion theory could not explain the findings of varying
local slope in the experiments. Zhao et al. proposed a stochastic growth
model based on bulk diffusion \cite{PhysRevE.60.4310}, which correctly
predicted the kinetic roughening phenomena observed in their
experiment \cite{ART_ZHAO_FORTIN_1} but lacks the details on
how polymers evolve. Insufficient theoretical studies coupled with
inconsistencies in the experiments motivate us to model the polymer film growth
and seek a better understanding of  the growth processes that determine the
roughening mechanism. In this paper, we study
the 1+1D lattice model for the  polymer films grown by VDP and examine the
effects of random angle deposition, free-monomer diffusion, free-monomer
adsorption  in determining the evolution of the film's morphology.
\begin{figure}
\centering
\includegraphics[width=0.9\columnwidth]{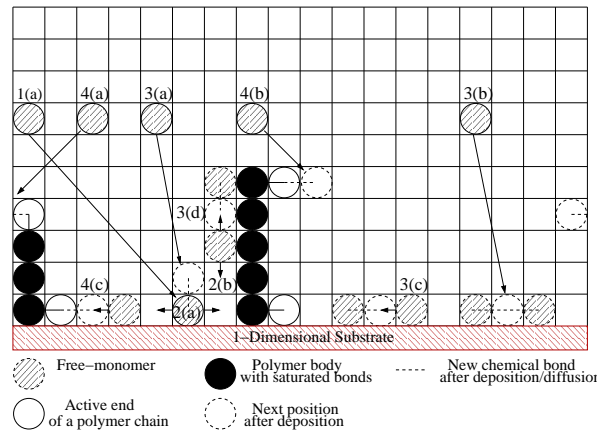}
\caption{ 
\label{fig_1_schematic}Schematic of the 1+1D growth model. 1(a): Free-monomer
deposition at random angles; 2(a): Adsorbed free-monomer diffuses along along
the substrate; 2(b): Adsorbed free-monomer diffuses on a polymer chain; 3(a,b):
Polymer chain initiation resulting from random angle deposition; 3(c,d): Polymer
chain initiation resulting from free-monomer diffusion on substrate and polymer
chain respectively; 4(a,b): Free-monomer deposits onto the active end of a
polymer chain resulting in chain propagation; 4(c) Chain propagation due to
free-monomer diffusion.
}
\end{figure}
\section{MODEL AND METHOD}
\indent
In our simulation,  the free-monomers were deposited at random angles on a 1-D
substrate of lattice size $L$  at a deposition rate $F$ (in units of monomers per site per unit time). 
The KISS random number generator \cite {Marsaglia_kiss_rng} was employed and one
deposition time unit corresponded to the deposition of $L$ free-monomers. The incident free-monomers were released from a
height  three lattice units above the highest point on the surface with an
initial abscissa  randomly chosen from $1\leq x \leq L$.  The depositing
free-monomers have a uniform ``launch angle'' distribution which corresponds to a
nonuniform flux distribution $J(\theta)\sim1/cos(\theta)$ of particles above the
surface, where $\theta $ is defined as the angle between the direction of
impinging monomer and the substrate normal \cite{PhysRevE.66.021603,PhysRevB.62.2118}. Our VDP growth model is similar to the square-lattice disk
model studied by Ref. \cite{PhysRevE.66.021603} along with additional constraints of free-monomer diffusion, limited bonding, polymer-initiation,
propagation and merger. The particles followed a ballistic
trajectory until contacting the surface. The impinging free-monomer was then
moved to the lattice position nearest to the point of contact. The deposited
free-monomers were allowed to diffuse via nearest-neighbor hops with a diffusion
rate $D$ (nearest neighbor hops per monomer per unit time). A free-monomer that
is deposited on top of an existing polymer chain gets adsorbed on the chain and
is also allowed to diffuse. The excluded volume
constraint was implemented by rejecting the diffusion or deposition moves to an already occupied
site. As the monomer coverage increases on the substrate, the polymer film
grows along a direction perpendicular to the substrate. This ``two dimensional''
growth is often referred to as the  1+1D growth in  literature.
\begin{SCfigure*}
\centering
\includegraphics[width=1.65\columnwidth]{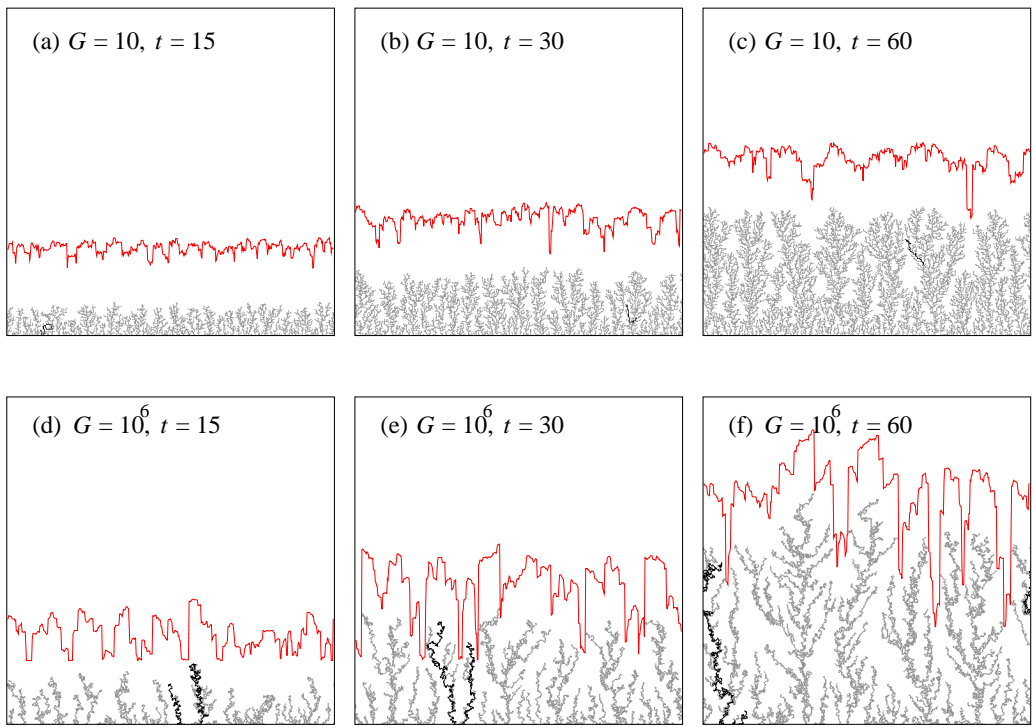}
\caption
{
\label{fig_2_six_eps_figs_grey}
Snapshots of the polymer films grown on a $L$ = $512$ substrate. Figs. (a, b, c)
were obtained for $G$ = 10 at deposition times $t$ = 15, 30, and 60 units respectively;
whereas Figs. (d, e, f) were obtained for $G$ =  $10^6$ and deposition times $t$
= 15, 30, and 60 units, respectively.  One deposition time unit corresponds to a
deposition of $L$ free-monomers.  For a clear view, the growth front (shown in red) is
displaced vertically by 100 pixels  and the longest polymer chain is
highlighted in black.
}
\end{SCfigure*}
\indent
Figure \ref{fig_1_schematic} shows the schematic of various processes that occur
during the non-equilibrium film growth on a 1-D substrate of length $L$ with 
periodic boundary conditions. Processes 1(a), 3(a), 3(b), 4(a), and 4(b) show
the gas phase free-monomers depositing onto the substrate at random  locations with uniform launch angle distribution. 
These free-monomers get adsorbed either on the substrate
(shown by process 2(a)) or on the polymer chains (process 2(b)). Adsorbed
free-monomers are allowed to diffuse along the adsorbent to any of the
nearest-neighboring unoccupied sites with equal probability, the rate of
diffusion $D$ is assumed to be the same on both the substrate  and the polymer
chain. We define  the chain length $s$ of a polymer as the number of monomers
forming a polymer chain. When an impinging free-monomer encounters another
free-monomer on the substrate as its nearest neighbor (process 3(a)), both  are
frozen and undergo a chemical reaction to form a dimer (polymer chain of length
$s=2$); polymers with chain length $s=3$ can also be formed after deposition
(process 3(b)).  Free-monomer diffusion along the substrate (process 3(c)) and
along the polymer chain (process 3(d))  can also result in the polymer chain
initiation \cite{BeachWilliamF._ma60061a014}. When an impinging free-monomer
encounters the active end of a polymer chain, it attaches itself to the chain
and increases the chain length by one unit (4(a), 4(b)). A diffusing
free-monomer can meet an active end of a polymer chain in its neighborhood and
get bonded to that polymer chain (process 4(c)). In linear polymer systems, the
free-monomers are allowed to form a maximum of two chemical bonds and at any
given time only the two ends of the polymer chain are chemically active,
resulting in the chain propagation at these two end locations. The chain portion
(of the polymer) excluding the two chemically active ends, is not allowed to
form chemical bonds with neighboring free-monomers. Free-monomers can however be
physically adsorbed on the chain and can diffuse along the chain (process 2(b)).
Another interesting process that occurs during the film growth is the
\textit{polymer-merger}. Two different polymers can merge when their respective
active-ends meet as nearest neighbors. During the polymer-merger, the nearest
neighboring ends of merging polymers react chemically to join the two polymers
into one longer polymer chain with higher molecular weight. The resulting
polymer chain is left with two active-ends, one from each of the parent
polymers. In the case when both the active-ends belonging to the same chain
appear as nearest neighbors, the chemical bond between the ends is prohibited.
In this study we do not attempt to study the effects of re-emission which allows
the free-monomers to ``bounce around'' before they settle at appropriate sites
on the surface. Instead, we assume that the impinging free-monomers will always
stick to the particle that comes on its deposition path (processes 4(a), 4(b)). 
\indent
At each stage of the simulation, either a deposition or a diffusion is
performed. In order to keep track of the competing rates of diffusion and
deposition we adapted the method suggested by Amar et al.\
\cite{PhysRevB.50.8781} and carried out the deposition with a probability
$p_{F}$,
\begin{equation}  
 \label{probability_deposition}
	p_{F}= \frac{1}{[1+N_{1}\times G ]} ,
\end{equation}
\noindent where $N_{1}$ is the free-monomer density (per site) and
$G=D/F$. The diffusion was carried out with the probability $p_{D}$,
\begin{equation} 
 \label{probability_diffusion}
	p_{D}= 1-p_{F}	= \frac{N_{1}\times G  }{[1+N_{1}\times G  ]} .
\end{equation}
\noindent
 In our simulations the incoming free-monomer flux $F$ was fixed for different
$D$, thus an increase in $D$ was parametrized as an increase in $G$.  Throughout
the growth process the  list of all free-monomers and polymer chains were continually updated. If a free-monomer encountered
another free-monomer or an active end of a polymer as its nearest neighbor, it
was added to the polymer chain and removed from the free-monomer list. In cases
where a free-monomer was the nearest neighbor to the active ends of more than two
polymers, we selected a random pair of polymers and performed polymer-merger.
\section{RESULTS}
\subsection{Surface Morphology}
In Fig. \ref{fig_2_six_eps_figs_grey} we show typical snapshots of the polymer
films generated using $L$ = $512$ substrate for two extreme cases: $G$ = $10$
(Figs. \ref{fig_2_six_eps_figs_grey}a, b, c) and $G$ = $10^6$ (Figs.
\ref{fig_2_six_eps_figs_grey}d, e, f)  after a deposition time of $t$ = 15
(Figs. \ref{fig_2_six_eps_figs_grey}a, d), $t$ = 30 (Figs.
\ref{fig_2_six_eps_figs_grey}b, e), and $t$ = 60 (Figs.
\ref{fig_2_six_eps_figs_grey}c, f) respectively. For both the values
of $G$, the films show the presence of  columnar structures, overhangs, and
unoccupied regions. These structures were observed to persist throughout the
growth process. Presence of these morphological structures can be explained by
the shadowing effect inherent in the growth process and is attributed to the 
cos$\theta$ distribution of the impinging free-monomer flux
\cite{PhysRevB.62.2118,PhysRevLett.62.788,PhysRevLett.63.692,
PhysRevLett.82.4882,PhysRevB.64.125411}. Shadowing effects arise when the
columnar structures  of
the surface ``stick out'' and shadow their neighboring sites, thus inhibiting
the growth in their neighboring sites. Due to the angular flux distribution of
the impinging free-monomers the taller surface features prevent the incoming
flux from entering the lower lying areas of the surface.

\indent
For comparable deposition times $t$, the films
grown at $G=10$ (Figs. \ref{fig_2_six_eps_figs_grey}a, b, c)  are
characterized by  small unoccupied-regions and short polymer chains, resulting
in shorter, denser, and compact films. Whereas for $G$ = $10^6$
(Figs. \ref{fig_2_six_eps_figs_grey}d, e, f), the films are characterized by
large unoccupied-regions and longer polymer chains resulting in taller,  more
porous, and less dense films. For a relatively low diffusion rate at $G$ = 10,
the free-monomers deposited on the film have a higher probability of
encountering another impinging free-monomer as nearest neighbor and thereby
initiating new polymers. Many such polymer initiations inhibit the occurrence of
unoccupied-regions and make the film dense and compact. In contrast, at a higher
diffusion rate of $G$ = $10^6$, the free-monomers have a higher probability to
diffuse upward towards the growth front and the upward diffusion of
free-monomers is favored due to the non-symmetric nature of the lattice
potential associated with diffusion over a step \cite{BOOK_BARABASI_STANLEY}.
The diffusing free-monomers arrive towards the growth front and bind to the
active ends of the polymers and increase their chain length. This explains the
occurrences of longer polymer chains at higher $G$ observed in Fig.
\ref{fig_2_six_eps_figs_grey}  (the longest polymer chains are highlighted in
black). Throughout the growth process the longest chains for $G$ = $10^6$ are much
longer than those obtained at $G = 10$. In Fig. \ref{fig_2_six_eps_figs_grey}
for  a fixed $t$ even though the films have the same number of particles, the
film morphology looks different for $G=10$ and $G=10^6$. The difference arises
from the variation in the position of growth front $h(x,t)$,  defined as the set of occupied sites in the film that are highest in each column, and 
$x$  represents the horizontal lattice site on the substrate. The growth front $h(x,t)$ studied here is a crude approximation of a more complex aggregate that is growing. However, our method of quantifying the growth front is justified because, in the AFM experiments, the measured 1-D height-height correlation function is based only on height profiles along the fast scanning axis. Moreover the finite size of the AFM tip is known to distort the growth front and the measured growth front is a convolution in which the interaction with the tip dilates the surface details \cite{aue:1347}. In Fig. \ref{fig_2_six_eps_figs_grey} the growth fronts  are shown vertically
displaced by $100$ pixels in the growth direction for clarity. The height fluctuation frequencies in $h(x,t)$ are observed to occur at different length
scales depending on the ratio $G$. For a fixed $t$, it is intuitive to think that the height
profiles corresponding to $G=10^6$ (Figs. \ref{fig_2_six_eps_figs_grey}d, e, f)
are more rougher than those obtained using $G=10$
(Figs. \ref{fig_2_six_eps_figs_grey}a, b, c).  A detailed analysis of the
dynamics of interface roughness evolution is presented in the later
section.
\begin{figure}
\centering
\includegraphics[width=0.9\columnwidth]{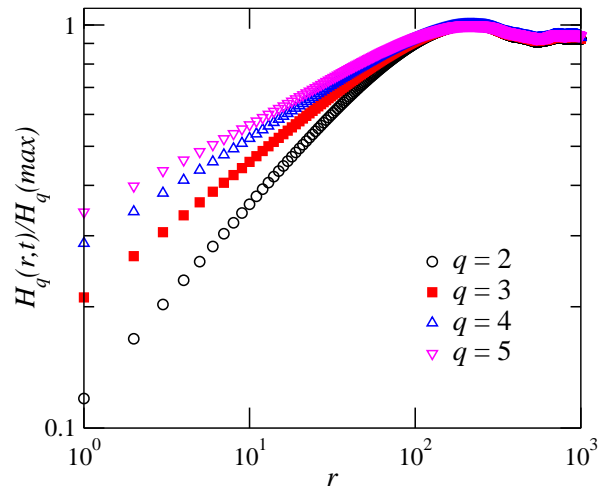}
\caption
{
	\label{fig_3_generalized_correlation_fns_G1_only} Generalized correlation function
$H_{q}(r,t)$ calculated for $L=2000$ and $t=1000$. The data are averaged over $200$ independent runs and the error-bars are smaller than the symbol size.
}
\end{figure}

\indent
The \textit{q}th-order generalized height-height correlation function $H_{q}(r,t)$ (defined below) is commonly used in identifying multi-affine surfaces  \cite{PhysRevA.44.2730,BOOK_BARABASI_STANLEY}.
\begin{equation}
\label{generalized_correlation_function}
	H_{q}(r,t) = \{\langle \mid h(x+r,t) - h(x,t)\mid^q\rangle\}^{(1/q)},
\end{equation}

\noindent
The scaling properties of \textit{multi-affine} surfaces can be described in terms of an infinite set of \textit{Hurst exponents} \cite{BOOK_MEAKIN,EDITED_FAMILY} $h_{q}$ which are obtained using,
\begin{equation}
\label{hurst_exponent}
	H_{q}(r,t)  \sim r^{h_{q} }.
\end{equation}

\noindent
For multi-affine interfaces,  the exponents $h_{q}$  are known to vary with $q$ \cite{BOOK_MEAKIN}. In Fig. \ref{fig_3_generalized_correlation_fns_G1_only}, we plot the generalized correlation function $H_{q}(r,t)$ using $q$ = $2$, $3$, $4$, and $5$ for $G$ = $10$ at deposition time $t$ = $10^3$. For smaller $r$, we observe a power-law dependence of $H_{q}(r,t)$ on $r$ in accordance to Eq. (\ref{hurst_exponent}). The slopes of the log-log plots shown in Fig. \ref{fig_3_generalized_correlation_fns_G1_only} depend on $q$ and indicate the presence of multi-scaling in the films generated by our VDP growth model. %Similar multi-affine interfaces were also observed for other studied $G$.

\subsection{Average Height and Growth Rate}
\begin{figure}
\centering
\includegraphics[width=0.9\columnwidth]{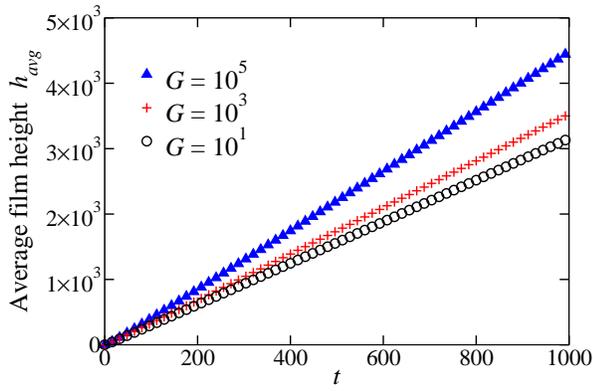}
\caption
{ 
\label{fig_4_avg_h_vs_G} Variation of $h_{avg}$ as a function of deposition time
$t$ for different $G$ and $L$ = 512. The data are averaged over 500 independent
runs. The statistical errors are smaller than the symbol size.
}
\end{figure}

\indent
The average height of the growth front $h_{avg}(t)$, is defined
as,
\begin{equation}
 \label{average_height}
	h_{avg}(t) = \dfrac{1}{L} \sum_{x=1}^L h(x,t),
\end{equation}

\noindent
and quantifies the overall thickness of the film. Figure \ref{fig_4_avg_h_vs_G} shows the plot of $h_{avg}$ versus  $t$ for different $G$ and
$L=512$. For the studied values of $G$, the
$h_{avg}$ versus $t$ plots show a linear relationship. This linear behavior is a
consequence of  restricting the growth to 1-D and the excluded
volume constraint implemented in the growth process. In Fig. \ref{fig_4_avg_h_vs_G} with an increase in $G$, we
observe a systematic  increase in the slope of the $h_{avg}$ versus $t$ plot. We define the \textit{growth rate} $R$ of the polymer
film as,
\begin{equation}
 \label{growth_rate}
	R = \frac{d h_{avg}}{dt}.
\end{equation}
\noindent
In general one expects the polymer film's growth rate to be proportional to the
incoming free-monomer flux only, i.e. $R$ $\propto$ $F$ and it is natural to
expect $R$ to be independent of $G(=D/F)$ since $F(=1/L)$ is a constant for a fixed $L$. However, in Fig. \ref{fig_4_avg_h_vs_G}
our simulations show a strong dependence of $R$ on $G$ as well, and the growth rate
$R$ is observed to increase monotonically with $G$.  This $G$ dependent growth rate $R$
indicates that the growth rate is effected by monomer diffusion directly, i.e.,
there is a net uphill monomer diffusion current that is responsible for this
$R-G$ relationship. We thus incorporate an additional term $r(G)$ that
determines $R$ in addition to its dependence on $F$.
\begin{figure}
\centering
\includegraphics[width=0.9\columnwidth]{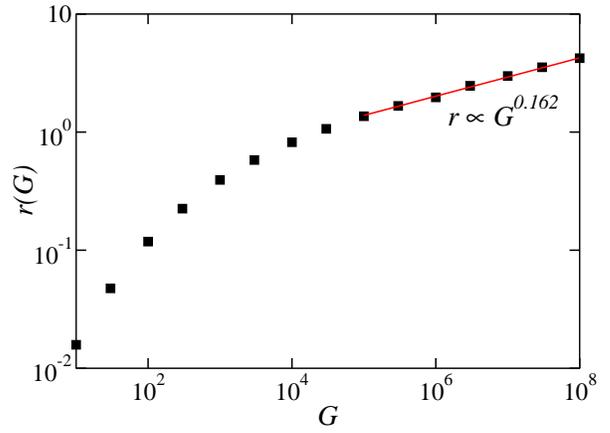}
\caption
{ 
\label{fig_5_r_vs_G} Variation of $r(G)$ as a function of $G$. The data were
obtained for $L$ = $512$ and averaged over $500$ independent runs. The statistical
errors are smaller than the symbol size.
}
\end{figure}
\begin{equation}
 \label{growth_rate_equation}
	R(G)=R_{0}+r(G),
\end{equation}
\noindent
where $R_{0}$ is the growth rate due to the random deposition flux only ($D$ =
0) and $r(G)$ is the growth rate induced only by the uphill diffusion of
free-monomers. In Fig. \ref{fig_5_r_vs_G} we show the variation of $r(G)$ as a
function of $G$. We find a monotonic increase in $r(G)$ with an increase in $G$.
This shows the strong influence of $G$ in determining the polymer film's growth
rate. Specifically, at higher $G$ we find $r(G)$ $\propto$ $G^{0.162(3)}$. This
indicates that the growth rate due to uphill diffusion $r(G)$ asymptotically
follows a power-law dependence with $G$.
\begin{figure}
\centering
\includegraphics[width=0.9\columnwidth]{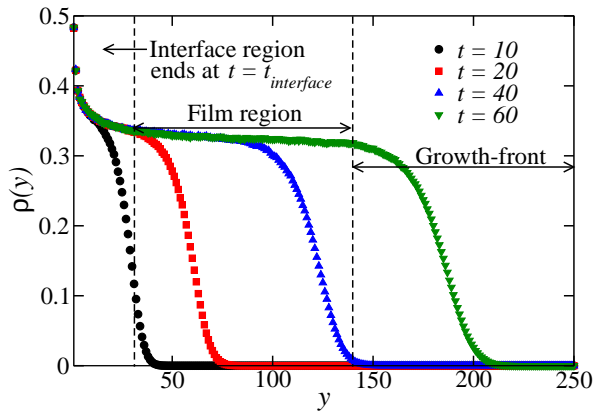}
\caption{
\label{fig_6_G1_L512_lateral_f_density_vs_t}
Lateral film density profiles $\rho \left( y\right)$ of polymer films grown on a
$L=512$ substrate with $G=10$ at deposition times of $t=10$, $20$, $40$,
and $60$ respectively. The statistical errors are smaller than the symbol size.
}
\end{figure}
\subsection{Characterization of the Kinetic Roughening}
\indent
The morphology of the growth front can be characterized by studying its
spatial and temporal evolution. In morphological scaling studies, typically two
kinds of scaling behaviors are associated with the roughening kinetics: the global
scaling and the local scaling \cite{refId2,PhysRevLett.72.2907,PhysRevLett.84.2199}.
The growth models  with anomalous kinetic roughening are known to have different scaling exponents
in the local and the global scales. The local roughness exponents have been employed in studying the irregularly growing mound morphology and are  often used in the experimental analysis \cite{ART_ZHAO_FORTIN_1,PhysRevLett.76.4931}. In general, the local roughness exponent $\alpha_{l}$ and the global roughness exponent $\alpha_{g}$ take different values depending on the type of scaling exhibited by the growth process. In the case of super-rough surfaces generated by nonequilibrium MBE growth models \cite{PhysRevE.49.122} and growth models with horizontal
diffusion \cite{PhysRevE.47.3242}, the assumption of the
equivalence between the global and local descriptions of the surface is not valid and such behavior has been termed as anomalous scaling
\cite{PhysRevE.49.122,0295-5075-24-7-010}. The differences in the global and local scaling exponents have been attributed to the
\textit{super-roughening} and \textit{intrinsically anomalous} spectrum observed in the anomalous scaling of surfaces \cite{PhysRevE.56.3993}.
\indent
For studying the morphological evolution of VDP generated polymer films, it is essential to identify their steady growth regime. To do so we employed
the \textit{lateral film density} $\rho \left( y \right)$ (at a height $y$ lattice units) of polymer film defined  as,
\begin{equation} 
 \label{lateral_density}
	\rho \left( y\right) =\frac{N\left(  y \right) }{L} ,
\end{equation}\noindent
where $N(y)$ represents the number of occupied lattice sites at a height
\textit{y} above the substrate. Calculations were performed on films with $L=512$,
$G=10$ at deposition times of $t=10$, $20$, $40$, and $60$,  respectively. Figure
\ref{fig_6_G1_L512_lateral_f_density_vs_t} shows the variation
of $\rho(y)$ with $t$. Similar plots were obtained for other $G$
values also. Three distinct regions of the polymer film growth referred to as
the \textit{interface region}, the steady-state \textit{film region} and the
\textit{growth front} were identified based on the Ref.
\cite{ART_BOWIE_ZHAO}. To evaluate the scaling exponents for kinetic roughening, we
only concentrate on films with the deposition time $t \gg t_{interface}$.
\begin{figure*}
\centering
\includegraphics[width=1.8\columnwidth]{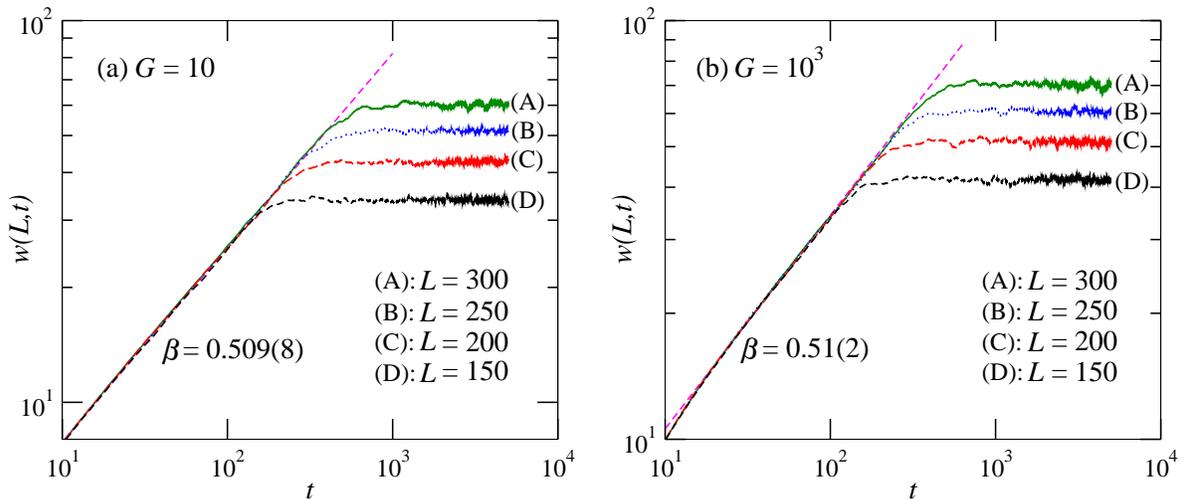}
\caption
{
\label{fig_7_combined_log_width_log_time_G1_G3} Time variation of the interface
width $w(L,t)$ for (a) $G = 10$ and (b) $G$ = $10^3$ with $L$ = $100$, $200$, and
$300$. The data are averaged over $1.8 \times 10^3$ independent runs.
}
\vspace{0.1in}
\end{figure*}

\subsubsection{Global Scaling Behavior}
\begin{figure}
\centering
\includegraphics[width=0.9\columnwidth]{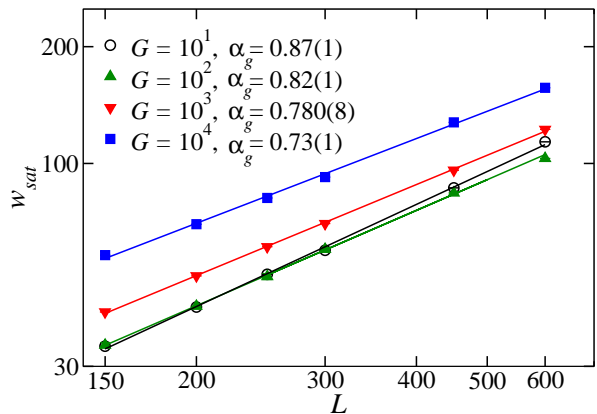}
\caption
{
	\label{fig_8_combined_alpha_G3_regression_using_L150_upto_L600}
Variation of $w_{sat}$ with $L$ for $G$ varying from $10$ to $10^4$. The 
data are averaged over  $1.8 \times 10^3$ independent runs and the statistical
errors are smaller than the symbol size.The error in the exponents were obtained from the curve-fit.
}
\end{figure}
The global scaling of the growth front can be determined by
studying the lattice size dependent \textit{interface width} $w(L,t)$ defined
as,
\begin{equation} 
\label{surface_width}
	w(L,t)=\sqrt{\dfrac{1}{L} \sum_{i}^L  \left[ h(i)- h_{avg} \right]  ^2}.
\end{equation}
In many 1-D morphological growth processes such as ballistic deposition
\cite{PhysRevA.38.3672,0305-4470-18-2-005}, Eden model
\cite{0305-4470-18-2-005}, and solid-on-solid models \cite{PhysRevLett.62.2289}, 
$w(L,t)$ usually follows the scaling law,
\begin{equation} 
\label{short_time}
	w(L,t)  \sim t^{\beta}	\qquad (t\ll t_{X}),
\end{equation}
\noindent
where exponent $\beta$ is known as the \textit{growth exponent} that
characterizes the time-dependence of  surface roughening. For any given $L$, the
power-law increase in $w(L,t)$ does 
\begin{figure*}
\centering
\includegraphics[width=1.8\columnwidth]{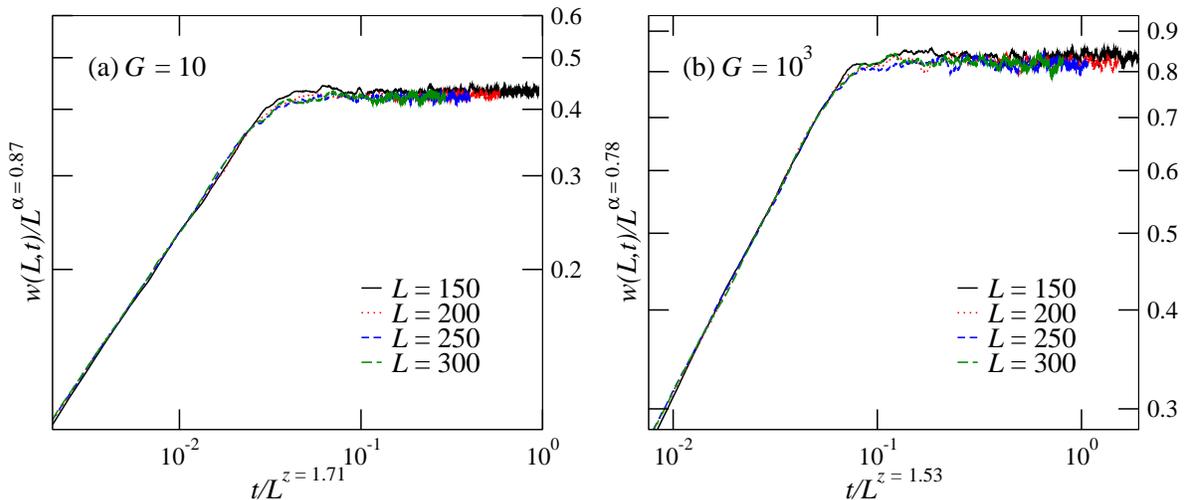}
\caption
{
\label{fig_9_combined_rescaled_plots_log_width_time}Rescaled plots of $w(L,t)/L^{\alpha_{g}}$ versus $t/L^{z_{g}}$.  The
``collapsed'' curves shown in the plots are the scaling functions for (a) $G=10$, (b) $G=10^3$ and indicate the presence on dynamic scaling. The data
are averaged over $1.8 \times 10^3$ independent runs.
}
\vspace{0.1in}
\end{figure*}
\noindent
not continue indefinitely with $t$, and is followed by a saturation regime during which the $w(L,t)$
reaches a saturation value $w_{sat}$. The power-law growth regime and the
saturation regime are separated by a \textit{crossover time} $t_{X}$. As $L$
increases, $w_{sat}$ also follows a power-law \cite{BOOK_BARABASI_STANLEY},
\begin{equation} 
\label{long_time}
	w(L,t) = w_{sat}\sim L^{\alpha_{g}}	\qquad (t\gg t_{X}),
\end{equation}
\noindent
and  $ {\alpha_{g}}$ is  referred to as the \textit{global roughness
exponent} \cite{BOOK_BARABASI_STANLEY}. The time $t_{X}$ at which the
behavior of $w(L,t)$ crosses over from Eq. (\ref{short_time}) to that of Eq.
(\ref{long_time}) depends on $L$ and scales as,
\begin{equation}
\label{time_crossover}
	t_{X} \sim  L^{z_g},
\end{equation}\noindent 
where $z_g$ is called the \textit{global dynamic exponent}. Typically the
scaling exponents are independent of specific interactions involved in the growth
process and depend on the dimensionality and symmetries of the system
\cite{0305-4470-18-2-005,0305-4470-19-8-006,BOOK_BARABASI_STANLEY,BOOK_MEAKIN}.
For some growth processes, the exponents $\alpha_{g}$, $\beta$, and $z_g$ are
unified using the \textit{dynamic scaling hypothesis} \cite{0305-4470-18-2-005} (also known as the \textit{Family-Vicsek scaling})  given as,
\begin{equation} 
\label{family_vicsek_relation}
	w(L,t) \sim L^{\alpha_{g}}\Psi\left( t/L^{z_g}\right),
\end{equation}\noindent
where $\Psi\left( t/L^{z_g}\right)$ is referred to as the \textit{scaling
function} and satisfies,
\begin{equation} 
\label{scaling_function_property} 
\Psi\left( x\right) = \left\{ \begin{array}{ll}
        x^{\beta} &	\qquad \qquad \left( x \ll 1\right)\\
        const &		\qquad \qquad \left( x \gg 1\right),\\
       \end{array}
       \right.
\end{equation}
\noindent and
\begin{equation}
 \label{defn_z}
z_g=\alpha_{g}/\beta.
\end{equation}
\indent
In Fig. \ref{fig_7_combined_log_width_log_time_G1_G3} we show two representative
plots of $w(L,t)$ versus $t$ on a log-log scale for (a) $G=10$, (b) $G=10^3$ and varying $L$.  From both Figs.
\ref{fig_7_combined_log_width_log_time_G1_G3}(a) and
\ref{fig_7_combined_log_width_log_time_G1_G3}(b), for $t\ll t_{X}$, the
$log(w(L,t))$ versus $log(t)$ shows a linear dependence implying a power-law behavior. For the same $G$ and $t \ll t_{X}$, the $w(L,t)$-$t$ plots overlap with one  another for different $L$. We estimated $\beta$ by fitting  Eq. (\ref{short_time})  to the plots of $w(L,t)$ in
Fig. \ref{fig_7_combined_log_width_log_time_G1_G3} for the film region $t \ll t_{X}$. We
obtained average $\beta=0.509(8)$ and $0.51(2)$ for $G$ = $10$ and $10^3$ respectively. For
other $G$ values, the $\beta$ obtained was close to $0.50$ and we observed an
invariance of  $\beta$ (within the error-bars) with $G$. The  statistical average and error bars in $\beta$ were
obtained from $1.8 \times10^3$ independent simulations.  The VDP model studied
here is similar to the ballistic deposition model with additional degree of freedom including diffusion, polymer initiation, extension and
merger. However, unlike the atomic
diffusion in MBE growth, the free-monomer diffusion is confined by the linear
geometry of the polymer chain. From Fig. \ref{fig_2_six_eps_figs_grey} one can
observe that in most cases the polymer chains are more or less perpendicular to
the substrate. Since the free-monomers can only move along the polymer chain,
most of the diffusion happens in vertical direction rather than in lateral
direction (which is the case for MBE growth). Yet, the vertical diffusion
does not contribute significantly to extra roughness increasing or decreasing
(the total particle number should be conserved) in the growth front due
to the porous nature of the film as compared to the random deposition model.
Thus, it is expected that the growth exponent will be close to that of
random deposition $(\beta \approx 0.50)$. We calculated $w_{sat}$ by  averaging $w(L,t)$ for \textit{$t \gg t_{X}$} from the data shown in Figs. \ref{fig_7_combined_log_width_log_time_G1_G3}(a) and
\ref{fig_7_combined_log_width_log_time_G1_G3}(b).  Figure \ref{fig_8_combined_alpha_G3_regression_using_L150_upto_L600} shows the plot of $w_{sat}$ versus
$L$ on the log-log scale  for  $G$ varying from $10$ to $10^4$. The exponent $\alpha_{g}$ was estimated for each $G$ using Eq. (\ref{long_time}). With an
increase in $G$ from $10$ to $10^4$, $\alpha_{g}$ was found to decrease from $0.87(1)$ to $0.73(1)$, the
error-bars in  $\alpha_{g}$ were obtained from the curve fitting. The
exponent $\alpha_{g}$ is known to be closely related to the surface fractal dimension
\cite{BOOK_BARABASI_STANLEY}. The smaller the $\alpha_{g}$, the larger is
the fractal dimension. Our observation of a decrease in $\alpha_{g}$ with an increase in $G$
shows that the fractal dimension of the growth front increases with $G$, i.e.
there are more spatial frequency fluctuations in the film morphology with
an increase in $G$. This finding is consistent with the growth front profiles shown in
Fig. \ref{fig_2_six_eps_figs_grey}. This result also demonstrates that  $G$
induces a large effective vertical growth rate $R(G)$ (shown in Fig.
\ref{fig_5_r_vs_G}) and the large  $R(G)$ in turn produces a much
rougher film surface.\indent \\ \indent
To determine whether the VDP process obeys the dynamic scaling behavior, we
start with an assumption that VDP growth follows the dynamic scaling hypothesis and
obtain $z_g$ through Eq. (\ref{defn_z}). As $w(L,t)$ scales with both $t$ (Eq.
(\ref{short_time})) and $L$ (Eq. (\ref{long_time})) we can rescale the $w(L,t)$
curve shown in Fig. \ref{fig_7_combined_log_width_log_time_G1_G3} by plotting
$w/L^{\alpha_{g}}$ versus $t/L^{z_g}$ to see whether those
curves ``collapse''. For  $G=10$ we obtain $\alpha_{g}$ =
$0.87(1)$, $\beta$ = $0.509(8)$ and according to Eq. (\ref{defn_z})  $z_g$ =
$1.71(1)$ and for $G=10^3$ we get  $\alpha_{g}$ = $0.780(8)$, $\beta$ =
$0.51(2)$, and $z_g$ = $1.53(2)$. We use the data from Figs.
\ref{fig_7_combined_log_width_log_time_G1_G3}(a) and
\ref{fig_7_combined_log_width_log_time_G1_G3}(b) and divide $w(L,t)$ by
$L^{\alpha_{g}}$. This shifts the curves of varying $L$ vertically on the log-log scale. According
to Eq. (\ref{long_time}), these curves now saturate at the same value of the
ordinate $w/L^{\alpha_{g}}$, however their
saturation times do not overlap. We then rescale the time axis and plot
$t/L^{z_g}$ for both cases of $G$. This rescaling of time axis
according to Eq. (\ref{time_crossover}) leads to a horizontal shift of the
curves and the curves now saturate at the same abscissa $t/L^{z_g}$.
In Figs. \ref{fig_9_combined_rescaled_plots_log_width_time}(a) and
\ref{fig_9_combined_rescaled_plots_log_width_time}(b) we show the rescaled plots of
$w(L,t)$ for $G=10$, $10^3$ and  consequently observe the ``collapse''
of individual curves for varying $L$ onto a single curve. This characteristic
``collapsed'' curve shown in  Figs.
\ref{fig_9_combined_rescaled_plots_log_width_time}(a) and
\ref{fig_9_combined_rescaled_plots_log_width_time}(b) is the \textit{scaling
function} $\Psi\left( t/L^{z_g}\right)$ mentioned in the Eq.
(\ref{scaling_function_property}). 
The scaling functions obtained in Figs.
\ref{fig_9_combined_rescaled_plots_log_width_time}(a) and
\ref{fig_9_combined_rescaled_plots_log_width_time}(b) are observed to follow Eq.
(\ref{scaling_function_property}) for both cases of $G = 10$ and $10^3$. For
other $G$ values, we observed similar ``collapse'' behavior indicating the global dynamic scaling of
growth fronts of the polymer films grown using VDP.
\begin{figure}
\centering
\includegraphics[width=0.9\columnwidth]{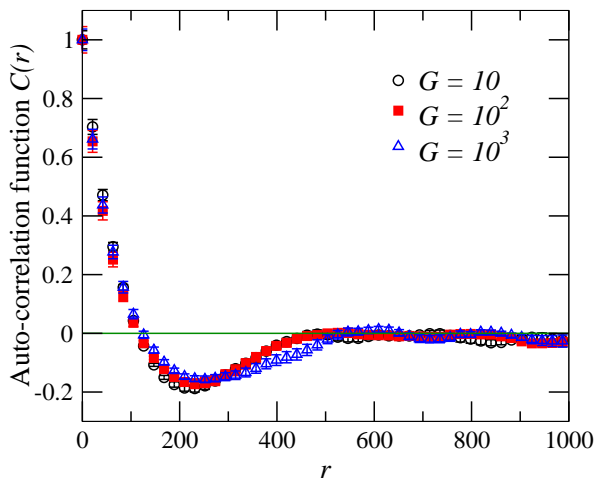}
\caption
{
	\label{fig_10_auto_correlation_fns_G1_only} Auto-correlation function
$C(r,t)$ calculated for $L=2000$ and $t=1000$ (averaged over $200$ independent runs).
}
\vspace{-0.25in}
\end{figure}
\begin{figure}
\centering
\includegraphics[width=0.9\columnwidth]{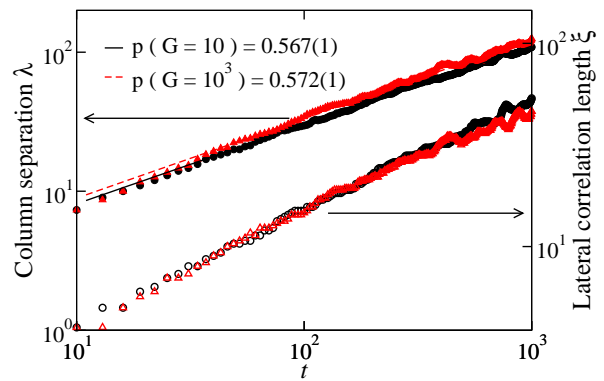}
\caption
{ 
\label{fig_11_feature_size_L2k_G1_3_t_1k} Estimates for column separation $\lambda$ and
lateral correlation length $\xi$ obtained for  $G=10 (circles)$ and  $G=10^3 (triangles)$. Open symbols correspond to  $ \xi $ and 
filled symbols correspond to $\lambda$. The data are averaged over $200$ independent
simulations with $L=2000$ and  $t=1000$. The error in the exponents were obtained from the curve-fit.
}
\end{figure}

\subsubsection{Local Scaling Behavior}
%below para from lopez and rodriguez in castro's paper
\indent
The local scaling behavior of the growth front can be understood from studying  the
spatial correlation functions: the auto-correlation function $C(r,t)$ and height-height
correlation function $H(r,t)$ that describe
the local properties of growing interfaces,
\begin{equation}
\label{C_r_equation}
	C(r,t) = \langle h(x+r,t) \cdot h(x,t) \rangle,
\end{equation}
\begin{equation}
\label{H_r_equation}
	H(r,t) = \{\langle [h(x+r,t) - h(x,t)]^2\rangle\} = 2[w^2-C(r,t)],
\end{equation}\noindent
where $r$ is the translation distance also referred to as the lag or slip
\cite{BOOK_ZHAO} and  $\langle...\rangle$  denotes a spatial average over the entire
system. The functions $C(r,t)$ and $H(r,t)$ are directly related as shown in Eq. (\ref{H_r_equation})  and differ
only by a constant pre-factor of $2w^2$. The $H(r,t)$ scales in the same way as
the interface width and is often used in studying the kinetic roughening
\cite{BOOK_BARABASI_STANLEY}. In order to obtain 
\begin{figure*}
\centering
\includegraphics[width=1.8\columnwidth]{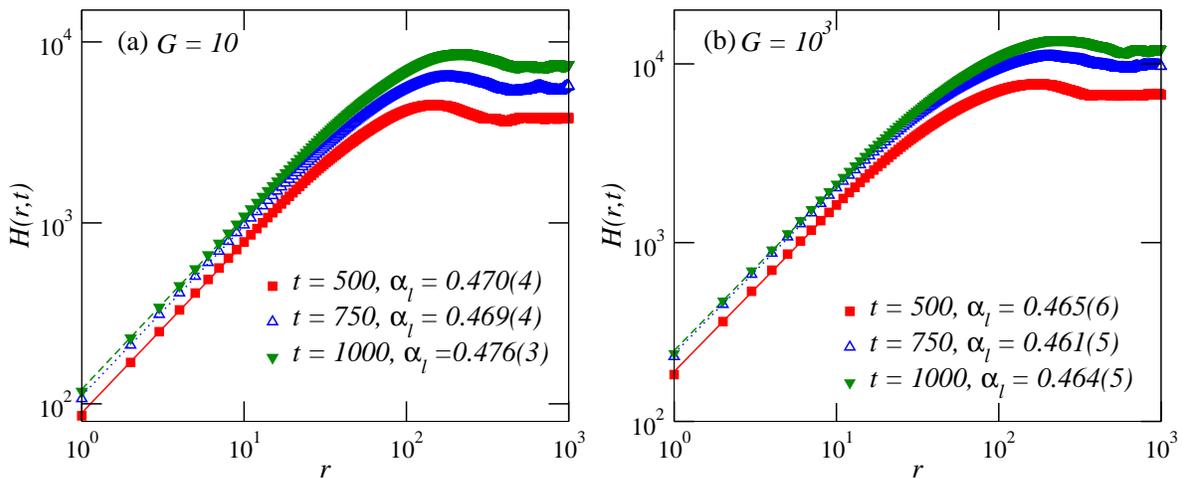}
\caption
{
	\label{fig_12_H_H_sq_L2k_G1_3_t_250_500_750_1k} Height-height correlation
function $H(r,t)$ for  (a): $G=10$, (b): $G=10^3$ using $L=2000$ at $t$ = $500$, $750$, and $1000$. The data are averaged
over  $200$ independent runs and error in the exponents were obtained from the curve-fit.
}
\end{figure*}
\noindent
accurate parameters from the correlation functions,
it is important to account for the accuracy of statistical averages.
\begin{figure*}
\centering
\includegraphics[width=1.8\columnwidth]{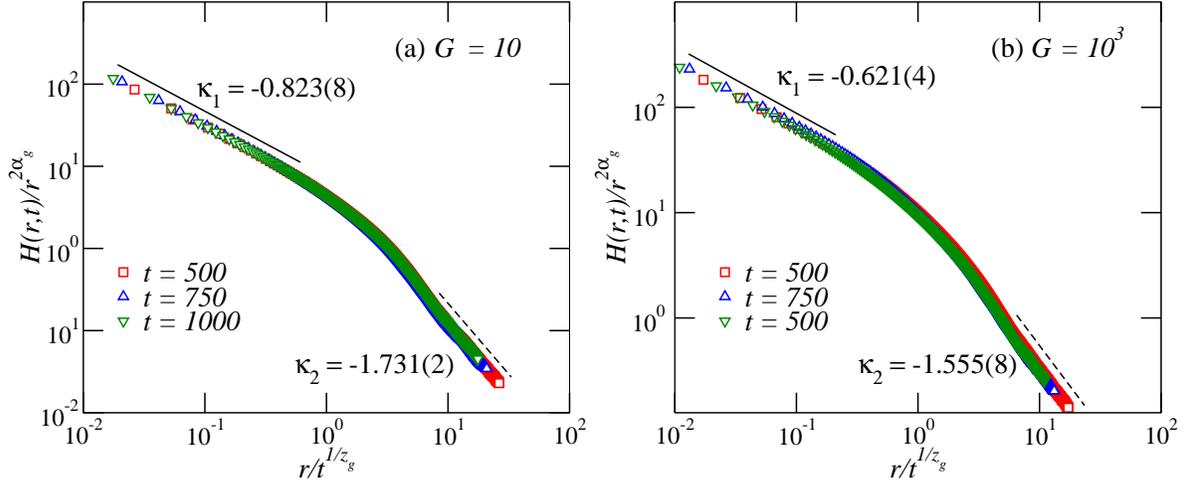}
\caption
{
	\label{fig_13_rescaled_H_H_sq_L2k_G1_3_t_250_500_750_1k} Plots of rescaled $H(r,t)$ showing the data collapse for  (a): $G=10$, and (b): $G=10^3$ using $L=2000$ at 
	$t$ = $500$, $750$, and $1000$. The data are averaged over  $200$ independent runs and error in the exponents were obtained from the curve-fit. The straight lines are plotted 
	as a guide to the eye. The scaling agrees with Eq. (\ref{g_A_scaling_function}).
}
\end{figure*}\noindent 
For the random Gaussian model surface
studied by Ref. \cite{BOOK_ZHAO}, converging $H(r,t)$ were obtained within an
order of $\left( \xi / L\right) ^{d/2}$ where $d$ is the system dimension and $\xi$ is the lateral correlation length. The
accuracy depends not only  on the number of data points,  but also on the sample
size. It is the ratio $\xi / L$, not the number of data points, that determines
the accuracy. Once the ratio $\xi / L$ is known, one may not be able to increase
the accuracy no matter how many data points are collected \cite{BOOK_ZHAO}. This
rule is different from the law of  large numbers for
independent random variables. This important difference needs to be recognized
while studying spatially correlated systems. Ideally, one would like to have  $\sqrt{ \xi / L} \ll 1$, i.e, $L \rightarrow\infty$.
However, due to the computational constraints we performed calculations using
$L=2000$ and $t=1000$. \indent
For a random self-affine surface, $C(r,t)$ usually decays to zero
with an increase in $r$. The shape of the decay depends on the type of the
surface and the decay rate depends on the distance over which two points $x$,
$x+r$ become uncorrelated. In Fig. \ref{fig_2_six_eps_figs_grey} the polymer
growth front does not appear to be a random rough surface, instead it has regular
fluctuations of the columnar structures. Figure \ref{fig_10_auto_correlation_fns_G1_only} shows the $C(r,t)$ for
$L=2000$ after a deposition time $t=1000$ for $G=10$, $10^2$, and $10^3$ used to characterize the surface morphology as a function of $t$. We
can define two different lateral length scales: the lateral correlation length $\xi$ and the average column-separation $\lambda$. The 
lateral correlation length $\xi$  defines a representative lateral dimension of a
rough surface and is estimated through $C(r,t)$ using $C(r=\xi,t)=C(0,t)/e$. Within a
distance of $\xi$  the surface heights of any two points are correlated. The parameter $\lambda$ characterizes a wavelength-selection of a surface and is determined by
measuring the value of $r$ corresponding to the first zero-crossing of $C(r,t)$ \cite{BOOK_ZHAO}. The variation of $\lambda$ with $t$ represents how the columnar
structures coarsen with deposition time. In general, the evolution of the columnar feature size follows a power law with $t$  given by \cite{ PhysRevE.66.021603}
\begin{equation}
\label{p_equation}
	\lambda \propto t^{p},
\end{equation}
\noindent
and  $p$ can be referred to as the \textit{coarsening exponent}.

\indent
In Fig. \ref{fig_11_feature_size_L2k_G1_3_t_1k} we plot 
$\lambda$ versus $t$ for $L=2000$ and $G=10$ and $10^3$ along
with the estimates for $p$. With an increase in $G$ from $10$ to $10^3$,   $p$ is found to remain close to $0.57$. 
The invariance of $p$  indicates that the coarsening of the columnar structures follow a power law that is unaffected by the ratio  $G$.
And we believe that this is dominated by the shadowing effect due to monomer vapor coming from all different angles uniformly.

\indent
The \textit{local roughness exponent} $ {\alpha_{l}}$ of the growth front  can be obtained from $H(r,t)$ using \cite{BOOK_BARABASI_STANLEY,refId2}
\begin{equation}
\label{local_alpha_equation}
	H(r,t) \sim r^{2 \alpha_{l}} \qquad (r \ll \xi).
\end{equation}\noindent
In Figs. \ref{fig_12_H_H_sq_L2k_G1_3_t_250_500_750_1k}(a) and   \ref{fig_12_H_H_sq_L2k_G1_3_t_250_500_750_1k}(b) we plot $H(r, t)$ for $t = 500$, $750$, and $1000$ along with the estimates  for ${\alpha}_{l}$ for $L=2000$ and  $G=10$, $10^3$ respectively. For a given $G$ and varying $t$, the estimates of ${\alpha_{l}}$ were observed to remain invariant within statistical-errors. From Figs. \ref{fig_12_H_H_sq_L2k_G1_3_t_250_500_750_1k}(a) and \ref{fig_12_H_H_sq_L2k_G1_3_t_250_500_750_1k}(b), we obtained an average  ${\alpha_{l}}$ of $0.470(3)$ and $0.460(3)$ for $G=10$ and $10^3$ respectively.   A comparison between $\alpha_{g}$ and $\alpha_{l}$ (shown in Fig. \ref{fig_8_combined_alpha_G3_regression_using_L150_upto_L600} and  Fig. \ref{fig_12_H_H_sq_L2k_G1_3_t_250_500_750_1k}) shows  $\alpha_{g}>\alpha_{l}$ for studied $G$ and the scaling behavior is observed to be different at short and large length scales. For small $r$, the plots of $H(r, t)$ do not overlap at varying coverage and show the presence of non-stationary anomalous scaling \cite{BOOK_MEAKIN}. The vertical temporal shift in the $H(r,t)$ observed in Fig \ref{fig_12_H_H_sq_L2k_G1_3_t_250_500_750_1k} is due to the difference between $\alpha_{g}$ and $\alpha_{l}$ and indicates the presence of anomalous scaling.  The mechanisms that lead to anomalous scaling can be separated into two classes: \textit{super-roughening} $({\alpha}_{g}>1)$ and \textit{intrinsic} anomalous scaling \cite{PhysRevE.54.R2189,Lpez1997329,PhysRevLett.84.2199}. The \textit{anomalous} growth exponent ${\beta}_{\ast} = ( {\alpha}_{g}-{\alpha}_{l} ) / z_g$ \cite{PhysRevE.56.3993} measures the difference between $\alpha_{g}$, $\alpha_{l}$ and can be obtained from the scaling of $H(r,t)$ \cite{Lpez1997329}
\begin{equation}
\label{rescaled_H_function}
	 H(r,t) = r^{2 \alpha_g} g_{A}( r/t^{1/z_g} ),
\end{equation}\noindent
and the  \textit{anomalous} scaling function $g_{A}(u)$  \cite{Lpez1997329,PhysRevE.56.3993} satisfies
\begin{equation}
\label{g_A_scaling_function}
g_A(u) \sim  \left\{ \begin{array}{ll}
        u^{-{\kappa}_1} &	\qquad \qquad \left( u \ll 1\right)\\
        u^{-{\kappa}_2} &	\qquad \qquad \left( u \gg 1\right)\\
       \end{array}
       \right.
\end{equation}
\begin{equation}
\label{equation_kappa_1}
  		{\kappa}_1 = 2( {\alpha}_{g}-{\alpha}_{l} ) \\
\end{equation}
\begin{equation}
\label{equation_kappa_2} 		
  		{\kappa}_2 = 2{\alpha_g}.
\end{equation}
\noindent
In Figs. \ref{fig_13_rescaled_H_H_sq_L2k_G1_3_t_250_500_750_1k}(a) and \ref{fig_13_rescaled_H_H_sq_L2k_G1_3_t_250_500_750_1k}(b)  we show the plots of ${H(r,t)}/r^{2\alpha_{g}}$ versus $r/t^{1/z_g}$ and obtain the ``data-collapse''  for $G=10$ and $10^3$ respectively.   The ``collapsed'' curves shown in  Figs. \ref{fig_13_rescaled_H_H_sq_L2k_G1_3_t_250_500_750_1k}(a) and \ref{fig_13_rescaled_H_H_sq_L2k_G1_3_t_250_500_750_1k}(b) are the scaling functions $g_{A}(u)$ (for $G=10$ and $10^3$) mentioned in Eq. (\ref{rescaled_H_function}). For both $G$, the scaling functions obtained in Fig. \ref{fig_13_rescaled_H_H_sq_L2k_G1_3_t_250_500_750_1k} satisfy  Eq. (\ref{g_A_scaling_function})  in accordance with  the theory for anomalous scaling \cite{Lpez1997329,PhysRevE.56.3993}.  The exponents $\kappa_1$ and $\kappa_2$ were obtained from  $g_{A}(u)$  plots using Eq. (\ref{g_A_scaling_function}).  For $G=10$ we have $\kappa_1 = 0.823(8)$,  $\kappa_2 = 1.731(2)$ and  for $G=10^3$ we obtained $\kappa_1 = 0.621(4)$, $\kappa_2 = 1.555(8)$. For both $G$, we find that  the exponents $\kappa_1$ and $\kappa_2$ obtained from the curve-fit satisfy Eqs.  (\ref{equation_kappa_1})  and (\ref{equation_kappa_2}) for the numerical estimates of $\alpha_g$ and $\alpha_l$ obtained from Eq. (\ref{long_time}) and Eq. (\ref{local_alpha_equation}).
\begin{figure}
\centering
\includegraphics[width=0.9\columnwidth]{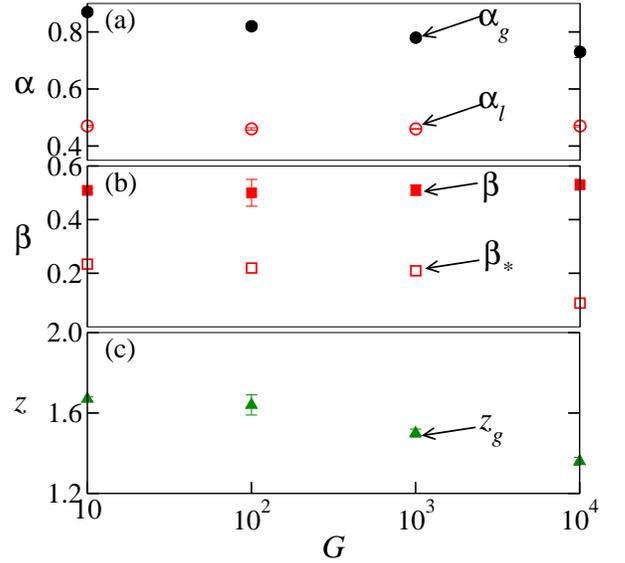}
\caption
{
	\label{fig_14_exponents_alpha_beta_z_VS_G} Variation of exponents
$\alpha_{g}$,  $\beta$, $z_g$, ${\alpha_{l}}$, and ${\beta}_{\ast}$ with $G$. 
The global exponents were calculated after  $t=5000$ and the data
were averaged over $1.8 \times 10^3$ independent runs. The local exponents
were calculated using $L=2000$ and $t=1000$. The data were
averaged over $200$  runs and the error in the exponents were obtained from the curve-fit.
}
%\vspace{-0.3in}
\end{figure}

\indent 
The dynamic scaling exponents of the kinetic roughening are  summarized in Figs. \ref{fig_14_exponents_alpha_beta_z_VS_G}(a), \ref{fig_14_exponents_alpha_beta_z_VS_G}(b), and \ref{fig_14_exponents_alpha_beta_z_VS_G}(c)  which shows the variation of global scaling exponents $\alpha_{g}$,  $\beta$, $z_g$ and local exponents $\alpha_{l}$, ${\beta}_{\ast}$ with $G$. With an increase in $G$ from $10$ to $10^4$, we found $\beta \approx 0.50$, $\alpha_{g}$ decreased  from $0.87(1)$ to $0.73(1)$, and $z_g$  decreased from $1.71(1)$ to $1.38(2)$. On the local length scale, an increase in $G$ did not effect a noticeable change in $\alpha_{l}$ $(\alpha_{l} \approx 0.46)$ and ${\beta}_{\ast}$ was observed to decrease from $0.23(4)$ to $0.18(8)$. 
\section{Conclusions}
We have performed 1+1D Monte Carlo simulation of VDP growth process by
considering free-monomer deposition, free-monomer diffusion, polymer initiation,
extension and polymer merger.  The ratio $G$ of the free-monomer diffusion
coefficient $D$ to the deposition rate $F$ was found to have a strong influence on the film's
growth morphology. The growth rate $R(G)$ of the polymer film was found to increase
monotonically with an increase in $G$. This is due to the consequence of an upper diffusion flux of free-monomers. The detailed analysis
of the surface morphology indicated the presence of very different scaling behavior at global and
local length scales. The kinetic roughening study of film interface indicates anomalous scaling and multiscaling. With an increase in $G$ from $10$ to $10^4$, the global growth exponent $\beta \approx 0.50$ was found to be invariant,  whereas the global roughness exponent $\alpha_{g}$ decreased from $0.87(1)$ to $0.73(1)$ along with a corresponding decrease in  the global dynamic exponent $z_g$ from $1.71(1)$ to $1.38(2)$.  The global scaling exponents were found to follow the dynamic scaling hypothesis with $z_g=\alpha_{g}/\beta$ for various $G$.  With an increase in $G$ from $10$ to $10^4$,  the average local roughness exponent $\alpha_{l}$ remained close to   $0.46$ with ${\alpha_{l}} \neq \alpha_{g}$,  this observation is unlike the ones obtained in self-affine surfaces \cite{EDITED_FAMILY,BOOK_BARABASI_STANLEY}.  The anomalous growth exponent ${\beta}_{\ast}$ was also found to  decrease from $0.23(4)$ to $0.18(8)$ with an increase in $G$.   Even though our model is in 1+1D as compared to the 2+1D experiments, our estimates of  $ \alpha_{l}$ and ${\beta}_{\ast}$ are close to the experimental findings of $\alpha \approx 0.5$ to $0.7$ and $\beta=0.25\pm 0.03$ obtained from the AFM studies of linear PA-N films grown by VDP \cite{ART_ZHAO_FORTIN_1, Streltsov}.  The similarity between the experimental and simulational estimates appears to be a coincidence since the dimension of the two systems are totally different.  We also did not observe the changes in the dynamic roughening behavior reported by Ref. \cite{lee:115427}, perhaps due to the limitations of our current simulation model in considering the effect of free-monomer diffusion only. This makes us believe that the kinetic roughening of the polymer films is sensitive to the specific molecular-level interactions, relaxations of polymer chains through inter-polymer interactions, and the intrinsic nature of polymerization process that need to be accounted for in the future simulations.

\section{ACKNOWLEDGEMENTS}
ST and DPL were partially supported by the NSF grant DMR-0810223. ST and YPZ
were also partially supported by NSF grant CMMI-0824728. ST would like to thank
S. J. Mitchell for help with the visualization tools.

%\bibliography{references}

\end{document}